\documentclass[12pt]{article}
\usepackage{graphicx}
\usepackage{dsfont}
\usepackage{cite}
\usepackage{amsmath}
\usepackage[T1]{fontenc}

\newcommand\pubnumber{}
\newcommand\pubdate{\today}

\def\institute{CERN Theory Department, 1211 Geneva 23, Switzerland\\
and\\
Nikhef Theory Group, Science Park 105, 1098 XG Amsterdam, The Netherlands}
\def\support{\footnote{Work supported by a Marie Slodowska-Curie Individual Fellowship of the
European Commission's Horizon 2020 Programme under contract number 704187.}}

\def\Title#1{\begin{center} {\Large #1 } \end{center}}
\def\Author#1{\begin{center}{ \sc #1} \end{center}}
\def\Address#1{\begin{center}{ \it #1} \end{center}}

\newcommand\pubblock{\rightline{\begin{tabular}{l} \pubnumber\\
         \pubdate  \end{tabular}}}
\newenvironment{Abstract}{\begin{quotation}  }{\end{quotation}}
\newenvironment{Presented}{\begin{quotation} \begin{center} 
             PRESENTED AT\end{center}\bigskip 
      \begin{center}\begin{large}}{\end{large}\end{center} \end{quotation}}
\def\Acknowledgements{\bigskip  \bigskip \begin{center} \begin{large}
             \bf ACKNOWLEDGEMENTS \end{large}\end{center}}

\def\beq{\begin{equation}}
\def\eeq#1{\label{#1}\end{equation}}
\def\eeqn{\end{equation}}

\def\beqa{\begin{eqnarray}}
\def\eeqa#1{\label{#1}\end{eqnarray}}
\def\eeqan{\end{eqnarray}}

\let\bar=\overbar

\def\Dslash{\not{\hbox{\kern-4pt $D$}}}
\def\dslash{\not{\hbox{\kern-2pt $\del$}}}

\def\msb{{\bar{\ssstyle M \kern -1pt S}}}

\begin{document}
\begin{titlepage}
\pubblock

\vfill
\Title{Top and Higgs: Recent Theory developments}
\vfill
\Author{Eleni Vryonidou\support}
\Address{\institute}
\vfill
\begin{Abstract}
In this talk I review recent theory developments in the computation of Higgs production in association with top quarks, as well as the modelling of the corresponding backgrounds. In addition to progress within the Standard model I discuss higher-order corrections for the $t\bar{t}H$ process in the presence of new interactions.
\end{Abstract}
\vfill
\begin{Presented}
$10^{th}$ International Workshop on Top Quark Physics\\
Braga, Portugal,  September 17--22, 2017
\end{Presented}
\vfill
\end{titlepage}
\def\thefootnote{\fnsymbol{footnote}}
\setcounter{footnote}{0}

\section{Introduction}
\vspace{-0.1cm}
The measurement of Higgs production in association with top quarks is expected to provide information about the top Yukawa coupling, which in turn can be a window to new physics. The LHC is already providing measurements of $t\bar{t}H$ production for different Higgs decay channels, and these are expected to become more precise as more luminosity is collected. Work is also needed on the theory side to provide accurate predictions. Next-to-leading order (NLO) QCD corrections for the $t\bar tH$ process have been computed in \cite{Beenakker:2002nc,Dawson:2003zu,Frederix:2011zi,Garzelli:2011vp},
with off-shell effects in \cite{Denner:2015yca}, whilst NLO electroweak (EW) results are given 
in \cite{Frixione:2015zaa,Hartanto:2015uka,Biedermann:2017yoi}. Recently results for resummation at  NLL \cite{Kulesza:2015vda,Broggio:2015lya} have been promoted to NNLL \cite{Kulesza:2017ukk,Broggio:2016lfj}. Finally, the combination of QCD and EW corrections including off-shell effects has been discussed in \cite{Denner:2016wet}. Significant progress has been made also related to the modelling of the $t\bar{t}bb$ background, in particular in assessing the uncertainties in relation to matching to the parton shower. 

Precise predictions for beyond the SM scenarios are equally
important at the LHC Run II. The Standard Model Effective Field Theory (SMEFT) provides a model-independent
framework to parametrise deviations from the SM via higher-dimension operators modifying the SM Lagrangian as follows: 
\begin{equation}  \mathcal{L}= \mathcal{L}_{SM}+\sum_i \frac{C_i}{\Lambda^2} O_i + \mathcal{O}(\Lambda^{-4}).
\end{equation}   
Fully differential NLO QCD corrections for top-quark production in the EFT have started to become
available. These include top-quark pair production, single top production and associated production \cite{Franzosi:2015osa,Zhang:2016omx,Bylund:2016phk,Maltoni:2016yxb}. QCD corrections can have a large impact on both the total cross sections and the differential distributions. NLO predictions come with  reduced theoretical uncertainties and can provide more reliable information for EFT fits. 

In this talk I focus on some recent developments in precise computations of the $t\bar{t}H$ process and the $t\bar{t}bb$ background, as well as the computation of higher order corrections for $t\bar{t}H$ in the presence of new interactions. 

\vspace{-0.3cm}
\section{Precision calculations for $t\bar{t}H$ and $t\bar{t}bb$}
\begin{figure}[h]
\begin{minipage}[h]{0.5\linewidth}
\centering
\includegraphics[trim=1cm 0.6cm 0 0, scale =0.8]{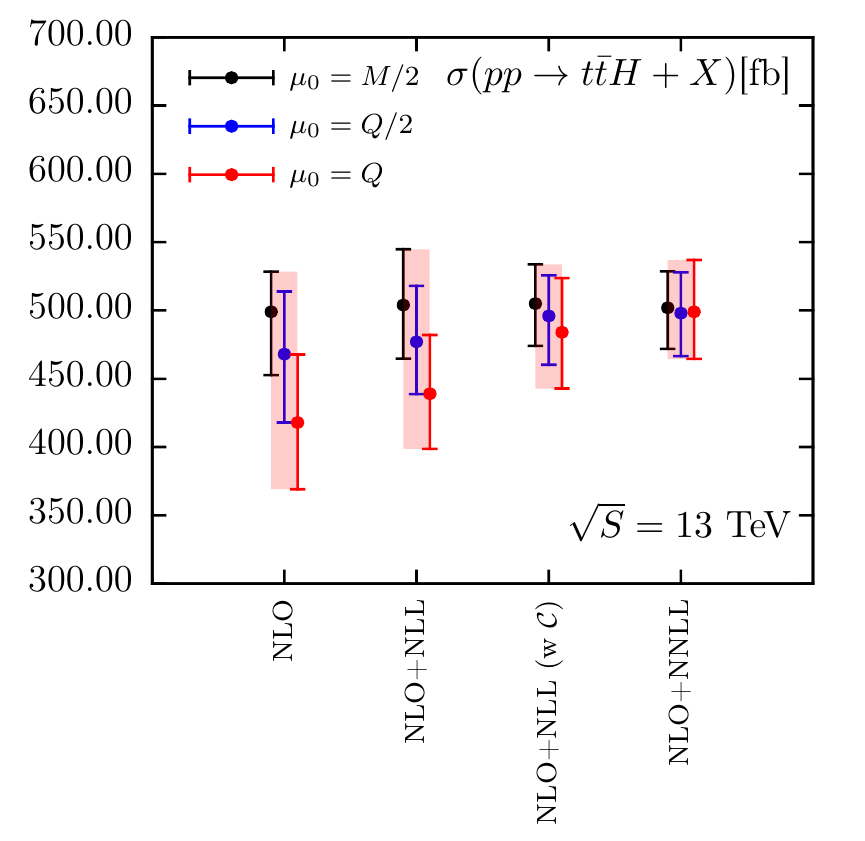}
\end{minipage}
\hspace{0.5cm}
\begin{minipage}[h]{0.5\linewidth}
\centering
\includegraphics[trim=5cm 0cm  2cm 3cm,scale=0.4]{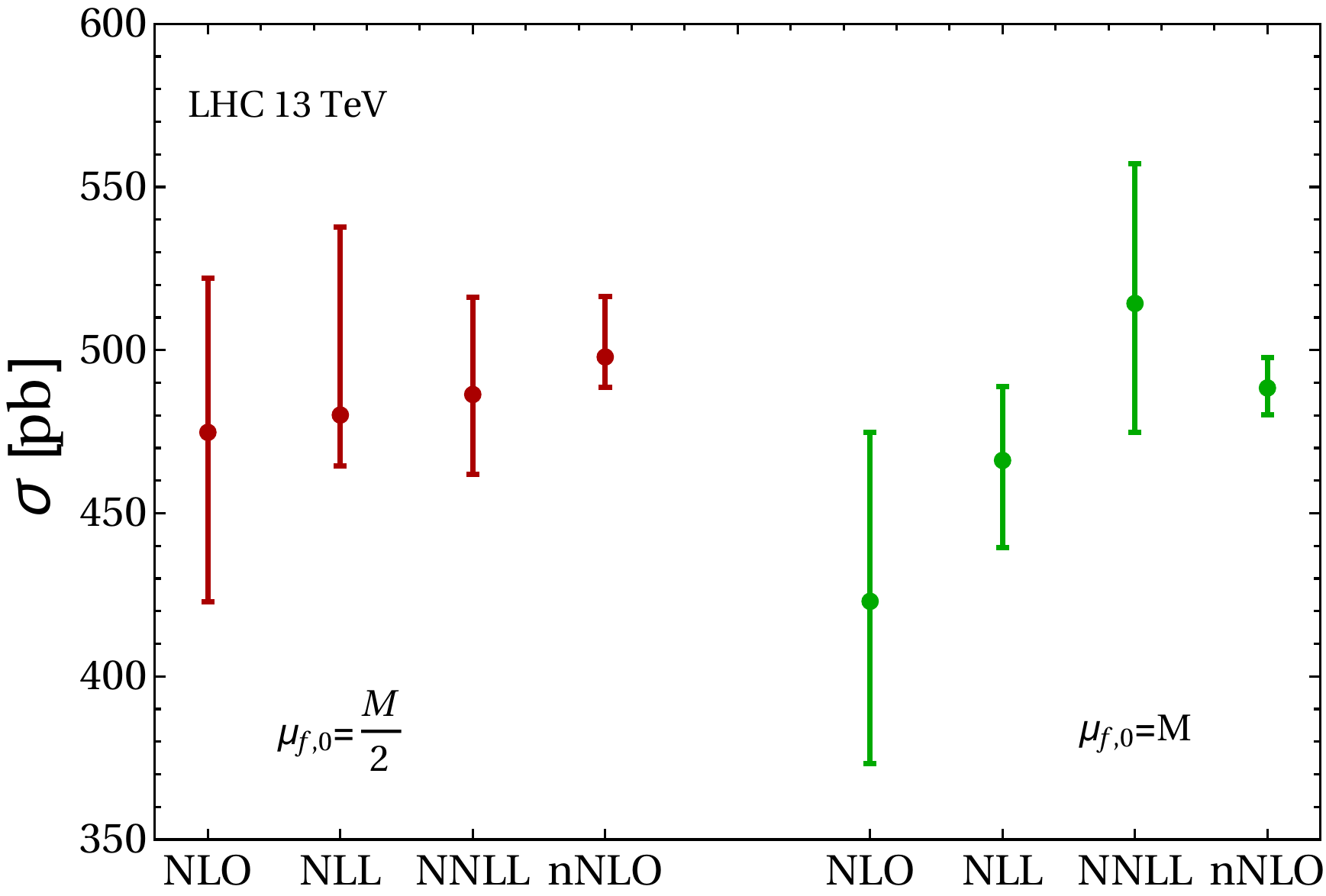}
\end{minipage}
\caption{\label{fig:resummation} Scale dependence of the $t\bar{t}H$ cross-section at 13 TeV at different orders for different scale choices. Left: results from \cite{Kulesza:2017ukk}. Right: results from \cite{Broggio:2016lfj}. For more details see the corresponding publications.} 
\end{figure}

Results for $t\bar{t}H$ production at NLO+NNLL have been obtained by two independent computations, one in the traditional resummation framework \cite{Kulesza:2017ukk} and one within SCET \cite{Broggio:2016lfj}. Both computations perform a soft gluon resummation, leading to improved predictions with reduced uncertainties as shown in fig. \ref{fig:resummation}. Differential results are also provided showing small changes in the distribution shapes, but a significant reduction of the scale uncertainties. 

\begin{figure}[h]
 \begin{minipage}[t]{0.5\linewidth}
\centering
\includegraphics[width=.85\linewidth,trim=1.5cm 0cm  0cm 0cm]{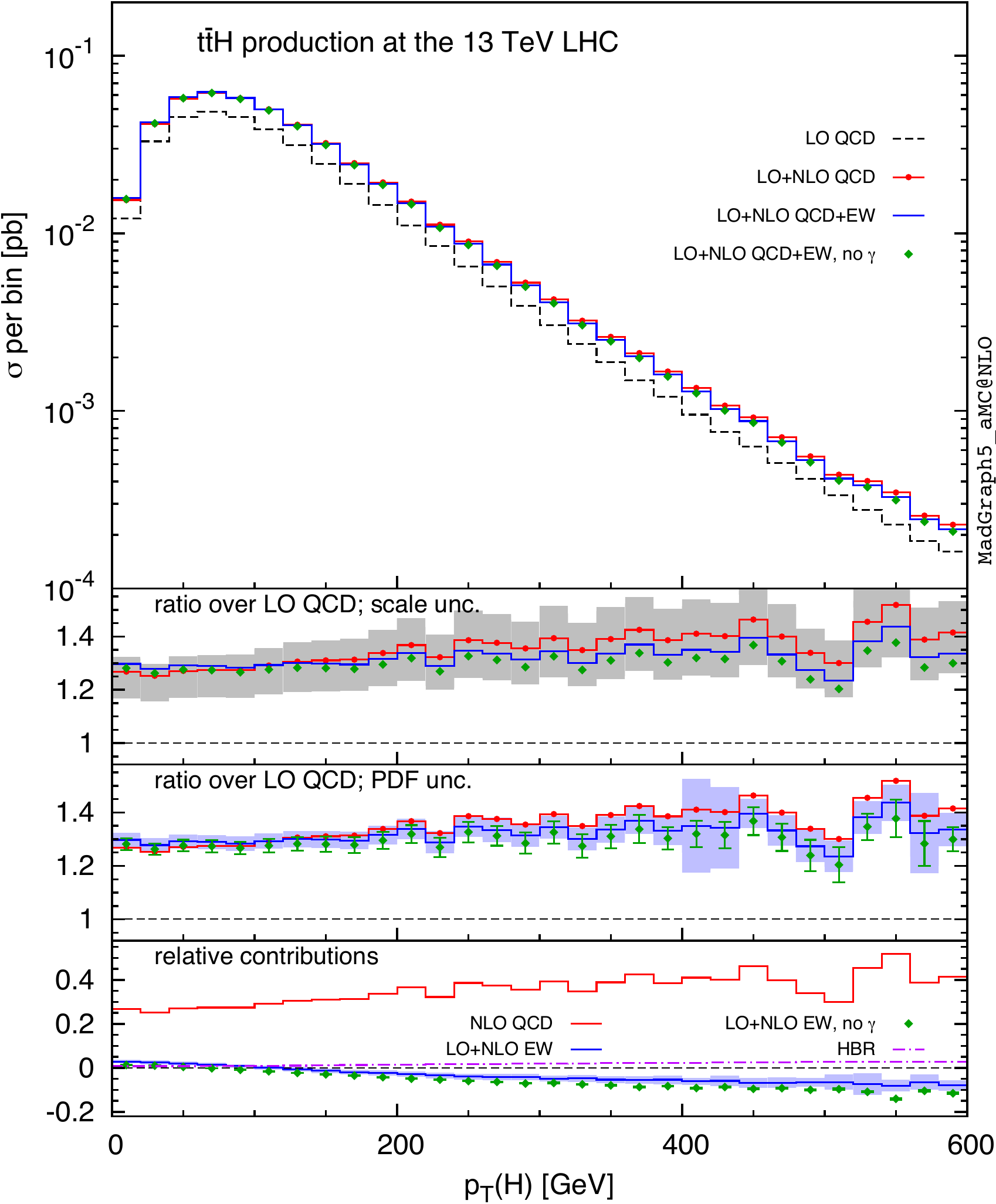}
\end{minipage}
\hspace{0.5cm}
 \begin{minipage}[b]{0.5\linewidth}
 \centering
 \includegraphics[trim=6cm 0cm  2cm 5cm,scale=0.65]{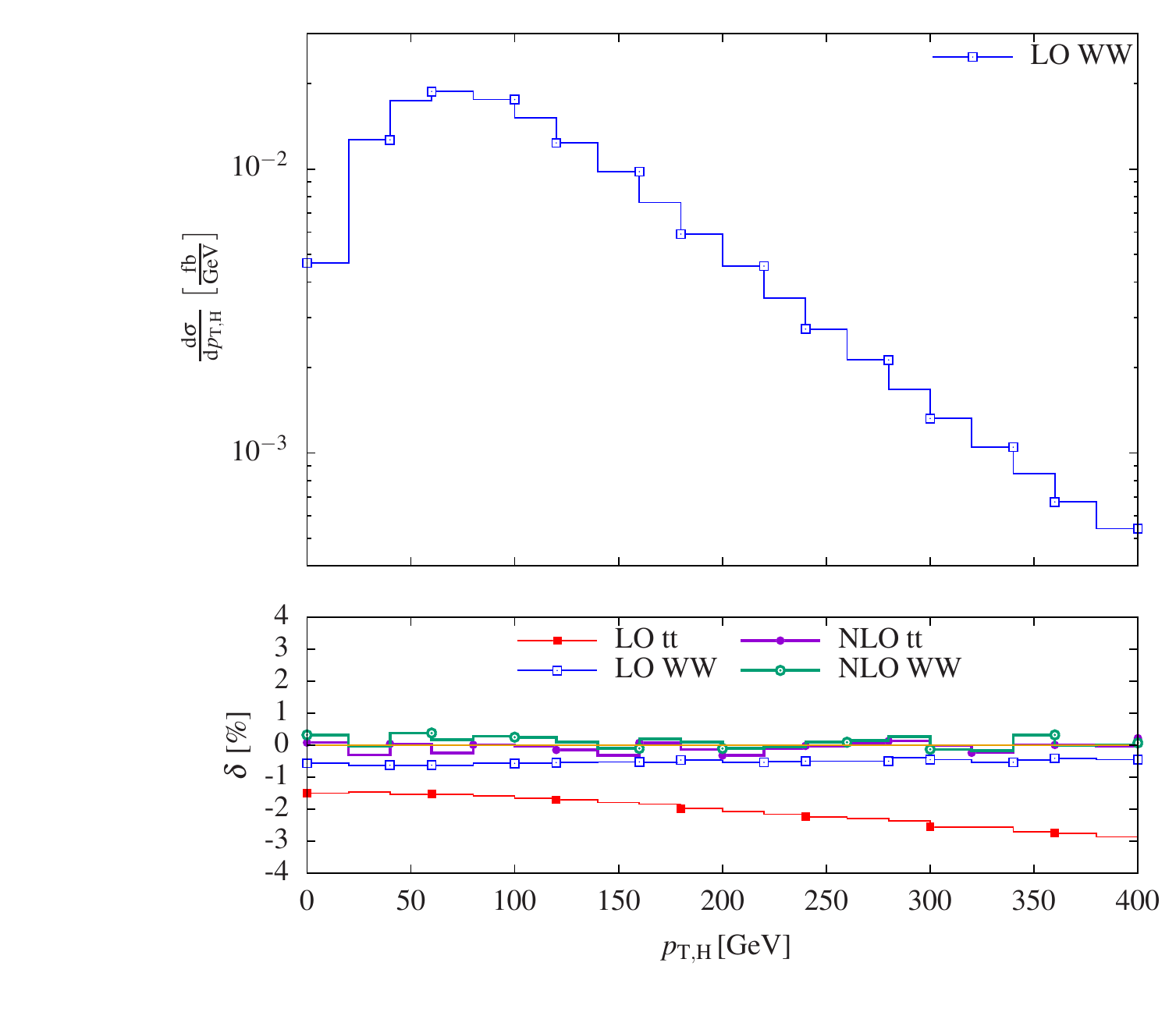}
 \end{minipage}
\caption{\label{fig:ptHew} 
Transverse momentum distribution of the Higgs boson in $t\bar{t}H$ at 13 TeV. Left: Impact of EW corrections from \cite{Frixione:2015zaa}. Right: Comparison with double pole approximation for top decays, taken from \cite{Denner:2016wet}. For more details see the corresponding publications.} 
\end{figure}

In addition to QCD corrections, significant progress has been made for EW corrections computed in \cite{Frixione:2015zaa,Hartanto:2015uka,Biedermann:2017yoi}. Whilst the impact of EW corrections is small at the total cross-section level, these are important in the high $p_T$ tails due to the Sudakov logs as shown in fig.~\ref{fig:ptHew} (left). Electroweak corrections have been also obtained for off-shell top quarks, taking also interference effects into account in \cite{Denner:2016wet}. These results are in agreement with the results for stable top quarks, and also show good agreement with the double pole approximation as shown in fig. \ref{fig:ptHew} (right). Finally a combination of EW and QCD corrections shows that additive and multiplicative combinations give almost identical results~\cite{Denner:2016wet}. 

\begin{figure}[h]
 \begin{minipage}[t]{0.5\linewidth}
\centering
\includegraphics[width=.97\linewidth,trim=0 0 1cm 0]{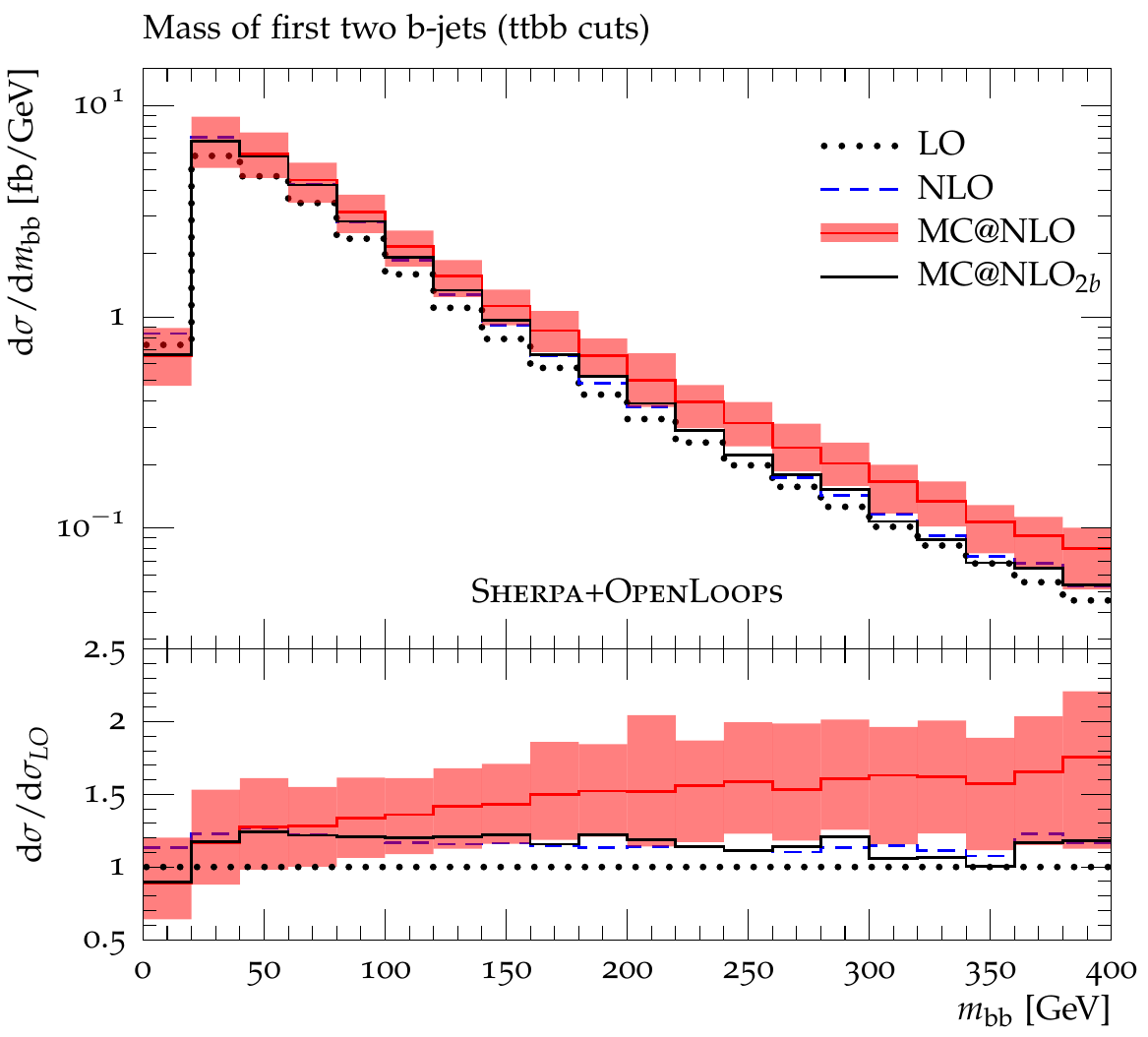}
\end{minipage}
\hspace{0.5cm}
 \begin{minipage}[b]{0.5\linewidth}
 \centering
 \includegraphics[width=.95\linewidth]{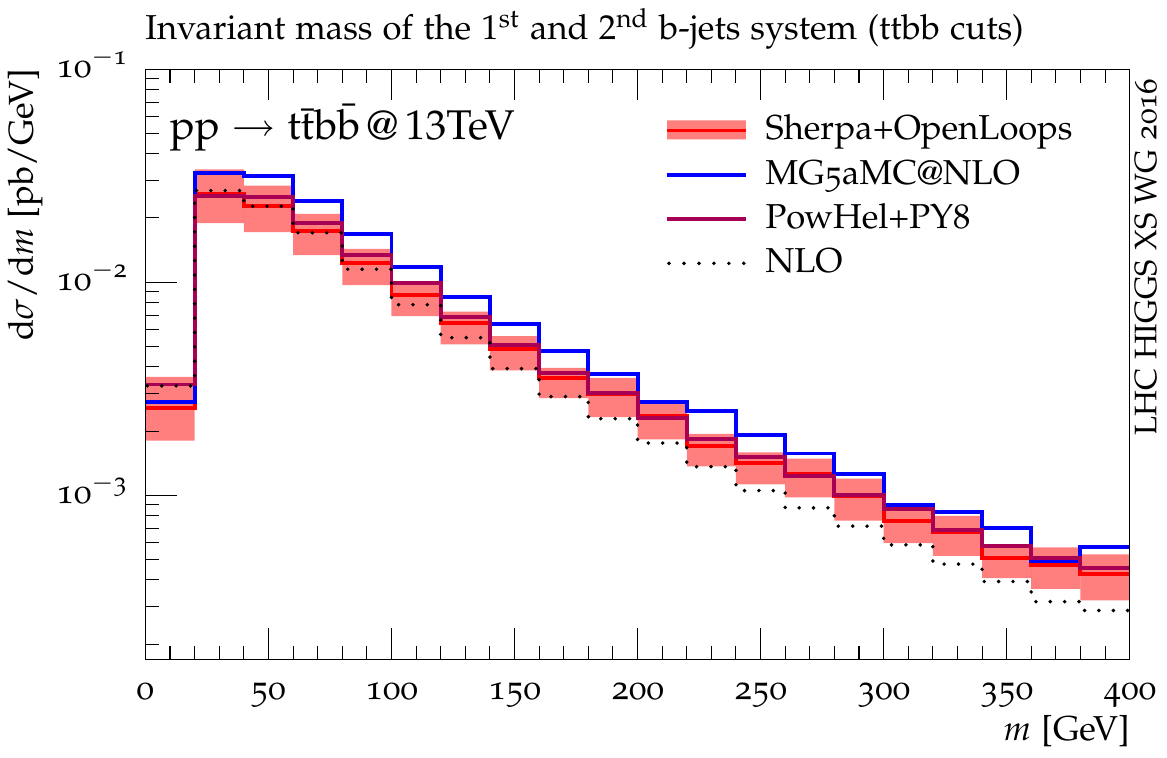}
  \includegraphics[width=.95\linewidth]{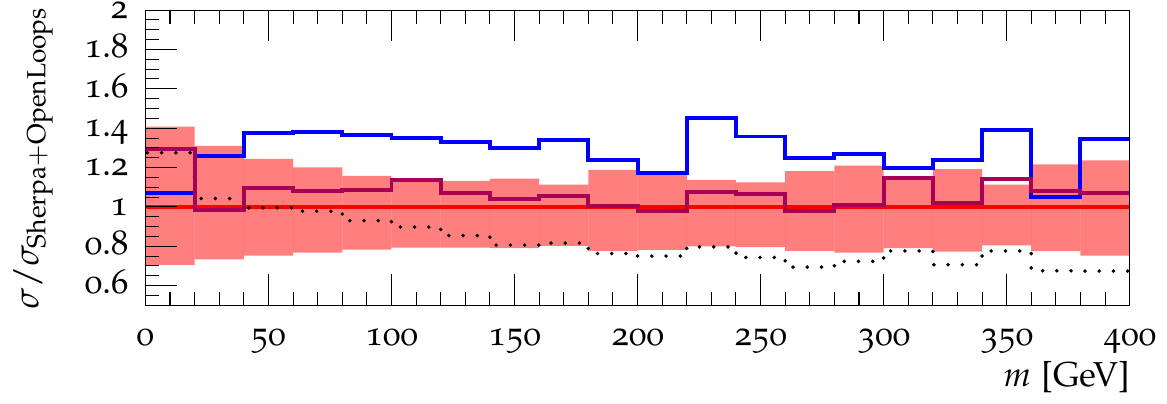}
 \end{minipage}
\caption{\label{fig:mbb} 
Invariant mass of the first two b-jets in $t\bar{t}bb$ at 13 TeV. Left: Comparison of NLO+PS to fixed order taken from \cite{Cascioli:2013era}. Right: Comparison of different generators and matching algorithms, taken from \cite{deFlorian:2016spz}.} 
\end{figure}

As measurements at the LHC improve in terms of statistics, it is also important to reduce the systematic uncertainties. One systematic uncertainty for $t\bar{t}H$ is the one coming from the modelling of the $t\bar{t}bb$ background. Several aspects enter the modelling of $t\bar{t}bb$ such as the use of the 4 or 5-flavour scheme, matching and merging of different jet multiplicities as well as parton-shower effects. The need to include the $t\bar{t}bb$ matrix elements to describe the process has motivated the computation of $t\bar{t}bb$ at NLO+PS in \cite{Cascioli:2013era}. The invariant mass distribution of the two hardest b-jets is shown in fig.~\ref{fig:mbb} (left). This study has shown that there is a large enhancement compared to the fixed-order results, due to secondary gluon splittings into $b\bar{b}$ pairs in the shower. These findings have also motivated a systematic comparison between Monte Carlo generators, as well as parton showers within the LHCHXSWG $t\bar{t}H$ subgroup \cite{deFlorian:2016spz}, as shown in fig.~\ref{fig:mbb} (right). Differences have been found between generators, with MG5\_aMC  \cite{Alwall:2014hca} showing a larger enhancement over the fixed order results compared to Sherpa+Openloops \cite{Cascioli:2013era} and Powhel \cite{Garzelli:2014aba}. To provide a better understanding of the differences, further comparisons are ongoing, including a new implementation in PowHeg and a detailed comparison of different parton shower codes with MC@NLO matching.

\section{$t\bar{t}H$ in the EFT}
\begin{figure}
\begin{minipage}[t]{0.5\linewidth}
 \renewcommand{\arraystretch}{1.8}
\newcommand{\xs}[7]{$#1^{+#2+#6+#4}_{-#3-#7-#5}$}
 \vspace{0pt}
 \scriptsize
\begin{tabular}{llllll}
		\hline
		13 TeV &$\sigma$ NLO &$K$-factors
		\\\hline
$\sigma_{SM}$& \xs{0.507}{0.030}{0.048}{0.007}{0.008}{0.000}{0.000}&1.09\\
$\sigma_{t\phi}$& \xs{-0.062}{0.006}{0.004}{0.001}{0.001}{0.001}{0.001}&1.13\\
$\sigma_{\phi G}$& \xs{0.872}{0.131}{0.123}{0.013}{0.016}{0.037}{0.035}&1.39\\
$\sigma_{tG}$&\xs{0.503}{0.025}{0.046}{0.007}{0.008}{0.001}{0.003}&1.07\\
$\sigma_{t\phi,t\phi} $&\xs{0.0019}{0.0001}{0.0002}{0.000}{0.000}{0.0001}{0.000}&1.17\\
$\sigma_{\phi G,\phi G} $&\xs{1.021}{0.204}{0.178}{0.024}{0.029}{0.096}{0.085}&1.58\\
$\sigma_{tG,tG}$& \xs{0.674}{0.036}{0.067}{0.016}{0.019}{0.004}{0.007}&1.04\\
$\sigma_{t\phi,\phi G}$ &\xs{-0.053}{0.008}{0.008}{0.001}{0.001}{0.003}{0.004}&1.42\\
$\sigma_{t\phi,tG}$& \xs{-0.031}{0.003}{0.002}{0.000}{0.000}{0.000}{0.000}&1.10\\
$\sigma_{\phi G,tG}$& \xs{0.859}{0.127}{0.126}{0.017}{0.022}{0.021}{0.020}&1.37\\
\hline
\label{table:tth}
\end{tabular}
\caption{NLO cross sections in pb for $ttH$ at 13 TeV and corresponding $K$-factors. Scale, EFT scale and 
PDF uncertainties are also included.}
 \end{minipage}	
  \hspace{0.5cm}
 \begin{minipage}[t]{0.5\linewidth}
  \vspace{0pt}
 \centering
\includegraphics[width=\linewidth, trim=0 0 2 0]{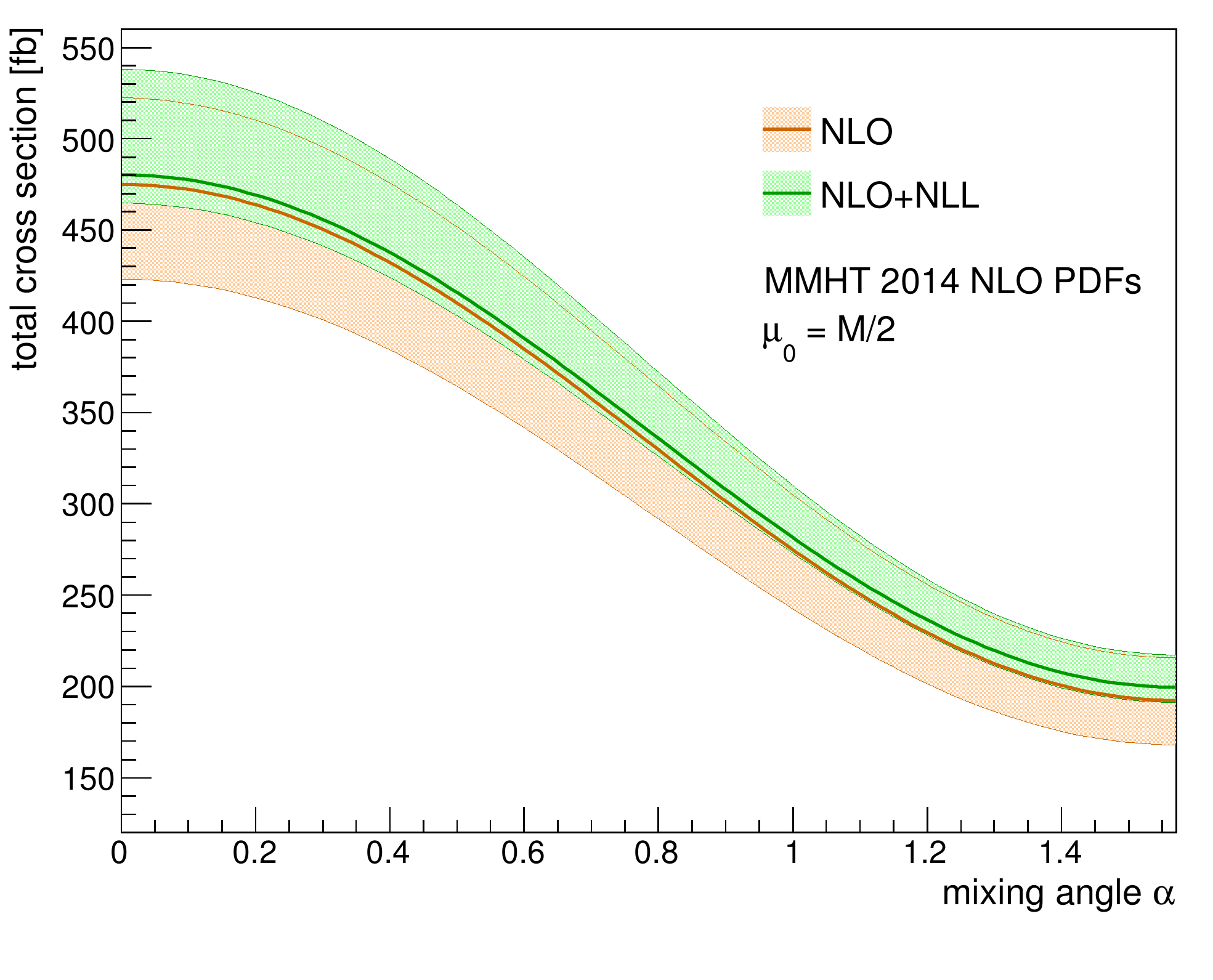}
\caption{Dependence of the $t\bar{t}H$ cross-section on the mixing angle at NLO+NLL from 
	\cite{Broggio:2017oyu} at 13 TeV with the corresponding scale uncertainties. 
	\label{fig:tthcp}}
\end{minipage}
\end{figure}
For $t\bar{t}H$ production the following operators contribute at dimension-6 \cite{Maltoni:2016yxb}:
\begin{eqnarray}\nonumber 
	&O_{t\phi} = y_t^3 \left( \phi^\dagger\phi \right)\left( \bar Qt \right)
	\tilde\phi \,,\,\,O_{\phi G} = y_t^2 \left( \phi^\dagger\phi \right) G_{\mu\nu}^A G^{A\mu\nu}\,,
        \\
	&O_{tG} = y_t g_s (\bar Q\sigma^{\mu\nu}T^A t)\tilde\phi G_{\mu\nu}^A\,.
	\label{operators}
\end{eqnarray}
 Four fermion operators and the triple gluon operator also contribute but these are expected to be constrained from top pair production and multijet processes respectively. 
The three operators above mix under RG flow, ~$O_{tG}$ mixes into $O_{\phi G}$, and both of them
mix into $O_{t \phi}$. Their contribution to the cross-section is parametrised as:
\begin{equation} 
	\sigma=\sigma_{SM}+\sum_i
	\frac{C_i}{(\Lambda/1\textrm{TeV})^2}\sigma_i^{(1)}+\sum_{i\le j}
	\frac{C_i C_j}{(\Lambda/1\textrm{TeV})^4}\sigma_{ij}^{(2)}\,.
	\label{sigma}
\end{equation}
Results for the cross sections at 13 TeV are summarised in fig.~4 using the notation of Eq.~\eqref{sigma}. The renormalisation and factorisation scale, 
EFT scale and PDF uncertainties are also shown. The EFT scale uncertainty is obtained by computing the EFT cross section at a different EFT scale taking mixing and running effects into account. A detailed discussion of the relevant uncertainties, RG effects and a comparison between full NLO  and RG corrections is presented in \cite{Maltoni:2016yxb}.

CP-violating couplings can also be searched for in $t\bar{t}H$ production. Ref. \cite{Broggio:2017oyu} computes the $t\bar{t}H$ cross-section in the presence of pseudoscalar couplings as described by: 
\begin{equation}
\mathcal{L}=-\dfrac{m_t}{v}\,\bar{\psi}(\textrm{cos}\,\alpha+i \,\textrm{sin}\,\alpha\,\,\gamma_5)\psi X_0.
\end{equation} The dependence of the cross section at NLO+NLL as a function of the mixing angle $\alpha$ is shown in fig. \ref{fig:tthcp}.

\begin{figure}[h]
 \begin{minipage}[t]{0.5\linewidth}
\centering
\includegraphics[width=.99\linewidth]{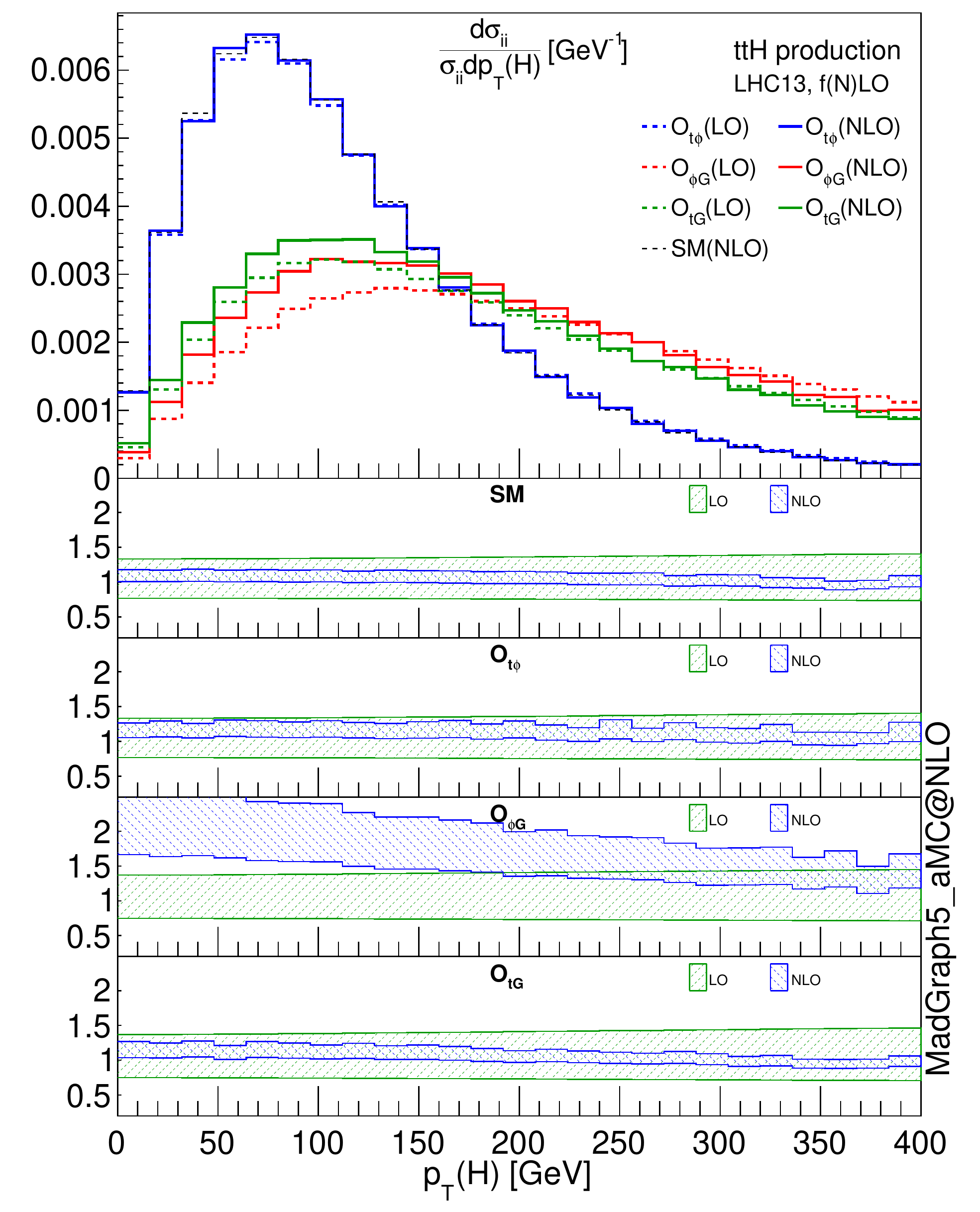}
\end{minipage}
\hspace{0.5cm}
 \begin{minipage}[b]{0.5\linewidth}
 \centering
 \includegraphics[width=.92\linewidth]{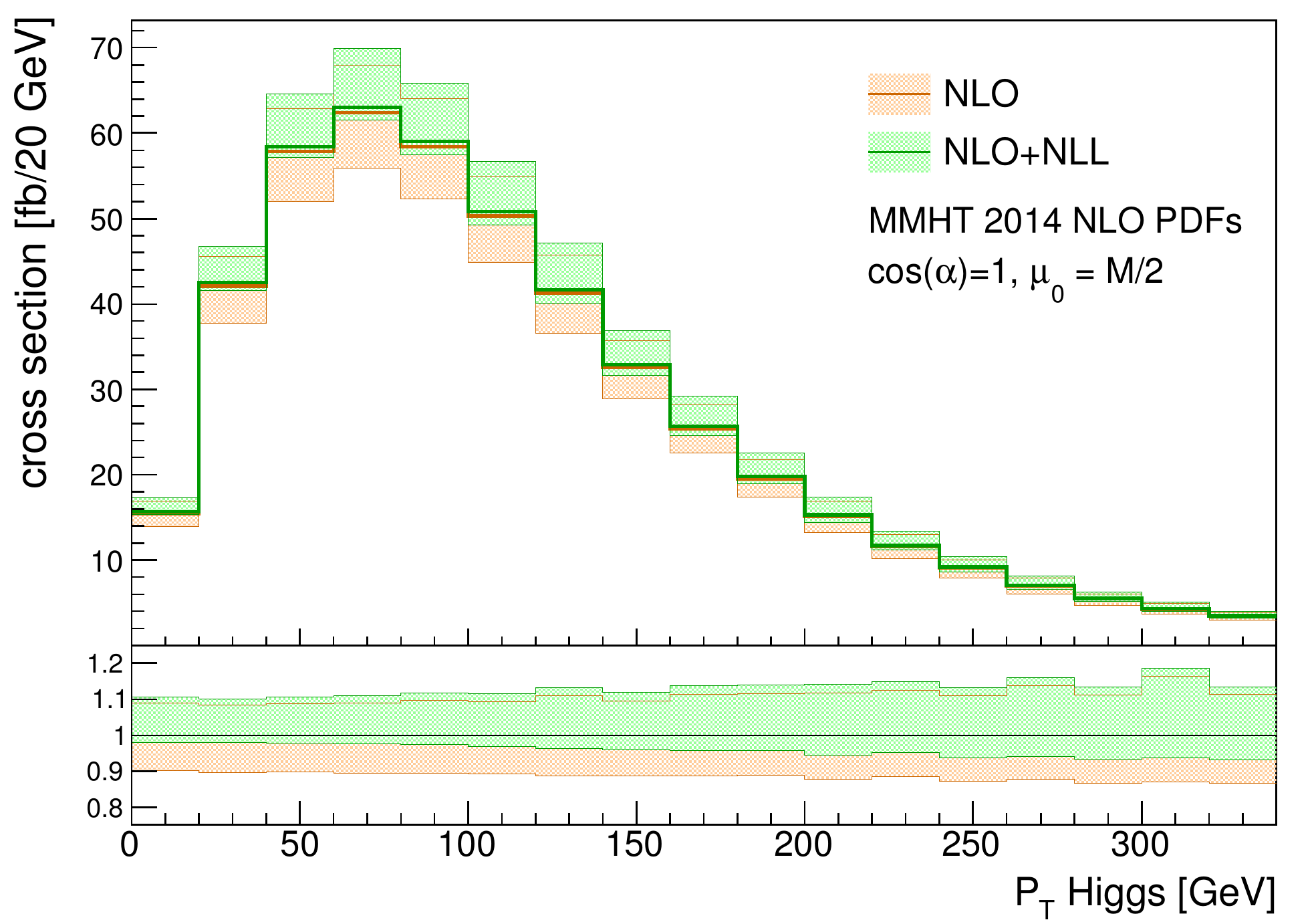}
  \includegraphics[width=.92\linewidth]{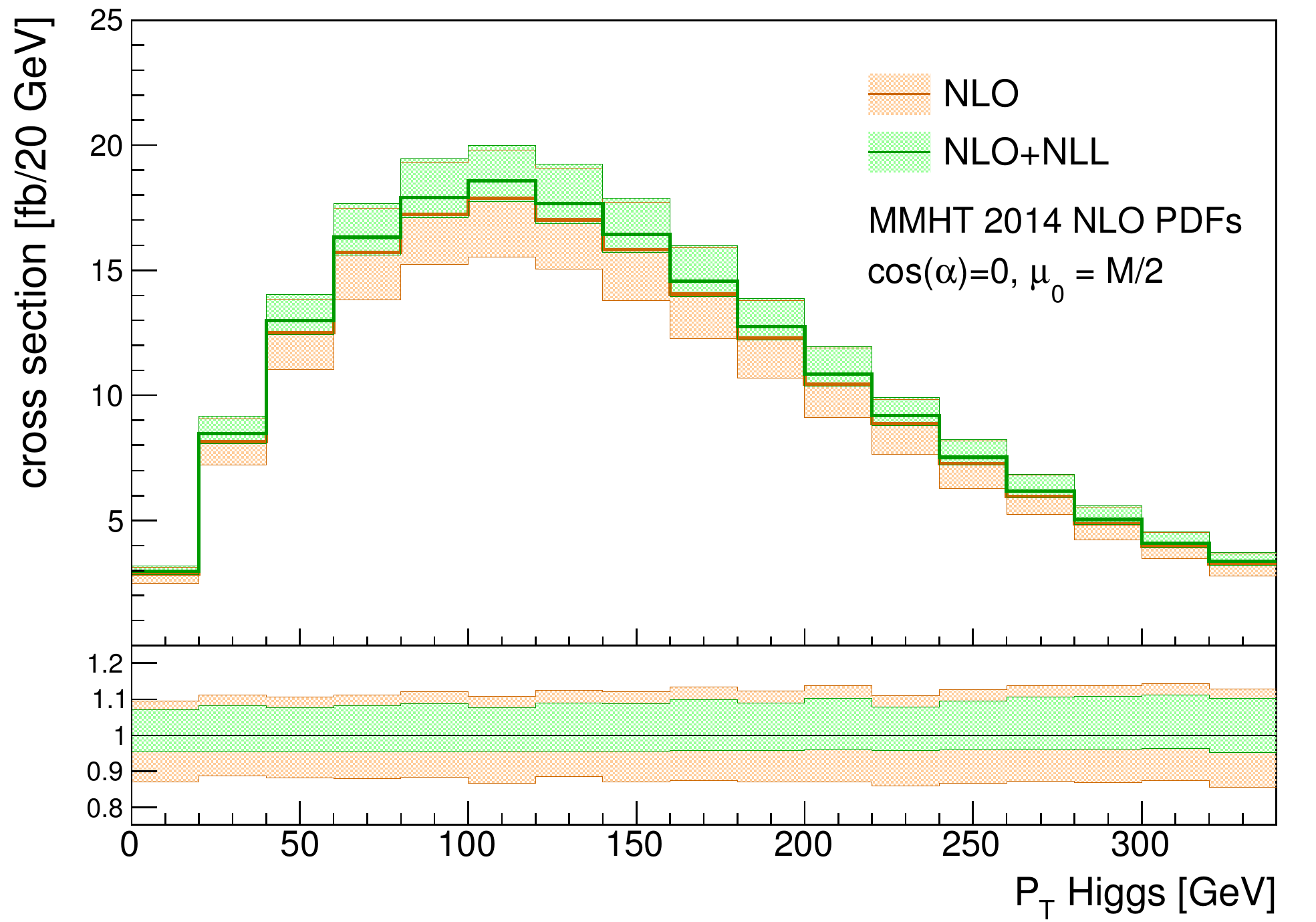}
 \end{minipage}
\caption{\label{fig:pt1} 
Transverse momentum distributions of the Higgs boson in $t\bar{t}H$ at 13 TeV. Left: SMEFT for different operators from \cite{Maltoni:2016yxb}. Lower panels give the $K$ factors and $\mu_{R,F}$ uncertainties. Right: distributions for the CP-even and CP-odd couplings from \cite{Broggio:2017oyu}. } 
\end{figure}

Differential distributions are also obtained with higher-order corrections.  Figure \ref{fig:pt1} (left) shows the $p_T$ of the Higgs in $t\bar{t}H$ at 13 TeV at LO and NLO (fixed-order) for the three operators of Eq.~\ref{operators}, demonstrating different shapes between different operators and non-flat $K$-factors~\cite{Maltoni:2016yxb}. Similarly the same distribution at NLO+NLL is shown for two different values of the mixing angle $\alpha$ in fig.~\ref{fig:pt1} (right) as obtained in \cite{Broggio:2017oyu}.

\section{Summary}
In these proceedings I summarised recent developments in the computation of $t\bar{t}H$ in the SM, including resummation, QCD and EW corrections. I reviewed also recent progress in the modelling of the $t\bar{t}bb$ background and progress in studying new top-Higgs interactions including higher order QCD corrections. 
\Acknowledgements
I would like to thank  F. Maltoni, M. Zaro and C. Zhang for their collaboration and useful discussions.


\begin{thebibliography}{99}

\bibitem{Beenakker:2002nc} 
  W.~Beenakker, S.~Dittmaier, M.~Kramer, B.~Plumper, M.~Spira and P.~M.~Zerwas,
  Nucl.\ Phys.\ B {\bf 653}, 151 (2003)
  doi:10.1016/S0550-3213(03)00044-0
  [hep-ph/0211352].


\bibitem{Dawson:2003zu} 
  S.~Dawson, C.~Jackson, L.~H.~Orr, L.~Reina and D.~Wackeroth,
  Phys.\ Rev.\ D {\bf 68}, 034022 (2003)
  doi:10.1103/PhysRevD.68.034022
  [hep-ph/0305087].


\bibitem{Frederix:2011zi} 
  R.~Frederix, S.~Frixione, V.~Hirschi, F.~Maltoni, R.~Pittau and P.~Torrielli,
  Phys.\ Lett.\ B {\bf 701}, 427 (2011)
  doi:10.1016/j.physletb.2011.06.012
  [arXiv:1104.5613 [hep-ph]].


\bibitem{Garzelli:2011vp} 
  M.~V.~Garzelli, A.~Kardos, C.~G.~Papadopoulos and Z.~Trocsanyi,
  EPL {\bf 96}, no. 1, 11001 (2011)
  doi:10.1209/0295-5075/96/11001
  [arXiv:1108.0387 [hep-ph]].


\bibitem{Denner:2015yca} 
  A.~Denner and R.~Feger,
  JHEP {\bf 1511}, 209 (2015)
  doi:10.1007/JHEP11(2015)209
  [arXiv:1506.07448 [hep-ph]].


\bibitem{Frixione:2015zaa} 
  S.~Frixione, V.~Hirschi, D.~Pagani, H.-S.~Shao and M.~Zaro,
  JHEP {\bf 1506}, 184 (2015)
  doi:10.1007/JHEP06(2015)184
  [arXiv:1504.03446 [hep-ph]].


\bibitem{Hartanto:2015uka} 
  H.~B.~Hartanto, B.~Jager, L.~Reina and D.~Wackeroth,
  Phys.\ Rev.\ D {\bf 91}, no. 9, 094003 (2015)
  doi:10.1103/PhysRevD.91.094003
  [arXiv:1501.04498 [hep-ph]].


\bibitem{Biedermann:2017yoi} 
  B.~Biedermann, S.~Bräuer, A.~Denner, M.~Pellen, S.~Schumann and J.~M.~Thompson,
  Eur.\ Phys.\ J.\ C {\bf 77}, 492 (2017)
  doi:10.1140/epjc/s10052-017-5054-8
  [arXiv:1704.05783 [hep-ph]].


\bibitem{Kulesza:2015vda} 
  A.~Kulesza, L.~Motyka, T.~Stebel and V.~Theeuwes,
  JHEP {\bf 1603}, 065 (2016)
  doi:10.1007/JHEP03(2016)065
  [arXiv:1509.02780 [hep-ph]].


\bibitem{Broggio:2015lya} 
  A.~Broggio, A.~Ferroglia, B.~D.~Pecjak, A.~Signer and L.~L.~Yang,
  JHEP {\bf 1603}, 124 (2016)
  doi:10.1007/JHEP03(2016)124
  [arXiv:1510.01914 [hep-ph]].


\bibitem{Kulesza:2017ukk} 
  A.~Kulesza, L.~Motyka, T.~Stebel and V.~Theeuwes,
  arXiv:1704.03363 [hep-ph].


\bibitem{Broggio:2016lfj} 
  A.~Broggio, A.~Ferroglia, B.~D.~Pecjak and L.~L.~Yang,
  JHEP {\bf 1702}, 126 (2017)
  doi:10.1007/JHEP02(2017)126
  [arXiv:1611.00049 [hep-ph]].


\bibitem{Denner:2016wet} 
  A.~Denner, J.~N.~Lang, M.~Pellen and S.~Uccirati,
  JHEP {\bf 1702}, 053 (2017)
  doi:10.1007/JHEP02(2017)053
  [arXiv:1612.07138 [hep-ph]].


\bibitem{Franzosi:2015osa} 
  D.~Buarque Franzosi and C.~Zhang,
  Phys.\ Rev.\ D {\bf 91}, no. 11, 114010 (2015)
  doi:10.1103/PhysRevD.91.114010
  [arXiv:1503.08841 [hep-ph]].


\bibitem{Zhang:2016omx} 
  C.~Zhang,
  Phys.\ Rev.\ Lett.\  {\bf 116}, no. 16, 162002 (2016)
  doi:10.1103/PhysRevLett.116.162002
  [arXiv:1601.06163 [hep-ph]].


\bibitem{Bylund:2016phk} 
  O.~Bessidskaia Bylund, F.~Maltoni, I.~Tsinikos, E.~Vryonidou and C.~Zhang,
  JHEP {\bf 1605}, 052 (2016)
  doi:10.1007/JHEP05(2016)052
  [arXiv:1601.08193 [hep-ph]].


\bibitem{Maltoni:2016yxb} 
  F.~Maltoni, E.~Vryonidou and C.~Zhang,
  JHEP {\bf 1610}, 123 (2016)
  doi:10.1007/JHEP10(2016)123
  [arXiv:1607.05330 [hep-ph]].


\bibitem{Cascioli:2013era} 
  F.~Cascioli, P.~Maierhöfer, N.~Moretti, S.~Pozzorini and F.~Siegert,
  Phys.\ Lett.\ B {\bf 734}, 210 (2014)
  doi:10.1016/j.physletb.2014.05.040
  [arXiv:1309.5912 [hep-ph]].


\bibitem{deFlorian:2016spz} 
  D.~de Florian {\it et al.} [LHC Higgs Cross Section Working Group],
  doi:10.23731/CYRM-2017-002
  arXiv:1610.07922 [hep-ph].


\bibitem{Alwall:2014hca} 
  J.~Alwall {\it et al.},
  JHEP {\bf 1407}, 079 (2014)
  doi:10.1007/JHEP07(2014)079
  [arXiv:1405.0301 [hep-ph]].

\bibitem{Garzelli:2014aba} 
  M.~V.~Garzelli, A.~Kardos and Z.~Trócsányi,
  JHEP {\bf 1503}, 083 (2015)
  doi:10.1007/JHEP03(2015)083
  [arXiv:1408.0266 [hep-ph]].

\bibitem{Broggio:2017oyu} 
  A.~Broggio, A.~Ferroglia, M.~C.~N.~Fiolhais and A.~Onofre,
  Phys.\ Rev.\ D {\bf 96}, no. 7, 073005 (2017)
  doi:10.1103/PhysRevD.96.073005
  [arXiv:1707.01803 [hep-ph]].












 
\end{thebibliography}
\end{document}